\documentclass[apj]{emulateapj}

\usepackage{color}
\usepackage{epsfig}
\usepackage{ulem}
\usepackage{graphicx}

\newcommand{\hMpc}{h^{-1}\rm{Mpc}}
\newcommand{\Ms}{h^{-1} M_{\odot}}

\begin{document}
\title{A First Look at creating mock catalogs with machine learning 
techniques}
\shorttitle{Making mocks with machine learning}
\author{
Xiaoying Xu\altaffilmark{1}, Shirley Ho\altaffilmark{1}, 
Hy Trac\altaffilmark{1}, Jeff Schneider\altaffilmark{1,2}, 
Barnabas Poczos \altaffilmark{2}, Michelle Ntampaka\altaffilmark{1}}

\begin{abstract}
We investigate machine learning (ML) techniques for predicting the
number of galaxies ($N_{gal}$) that occupy a halo, given the halo's
properties. These types of mappings are crucial for constructing the mock
galaxy catalogs necessary for analyses of large-scale structure. The
ML techniques proposed here distinguish themselves from traditional
halo occupation distribution (HOD) modeling as they do not assume a
prescribed relationship between halo properties and $N_{gal}$. In
addition, our ML approaches are only dependent on parent halo
properties (like HOD methods), which are advantageous over subhalo-based
approaches as identifying subhalos correctly is difficult. We test 2
algorithms: support vector machines (SVM) and k-nearest-neighbour (kNN)
regression. We take galaxies and halos from the Millennium simulation
and predict $N_{gal}$ by training our algorithms on the following 6
halo properties: number of particles, $M_{200}$, $\sigma_v$, $v_{max}$,
half-mass radius and spin. For Millennium, our predicted $N_{gal}$ values
have a mean-squared-error (MSE) of $\sim0.16$ for both SVM and kNN. Our
predictions match the overall distribution of halos reasonably well and
the galaxy correlation function at large scales to $\sim5-10\%$. In
addition, we demonstrate a feature selection algorithm to isolate
the halo parameters that are most predictive, a useful technique for
understanding the mapping between halo properties and $N_{gal}$. Lastly,
we investigate these ML-based approaches in making mock catalogs for
different galaxy subpopulations (e.g. blue, red, high $M_{star}$, low
$M_{star}$). Given its non-parametric nature as well as its powerful
predictive and feature selection capabilities, machine learning offers
an interesting alternative for creating mock catalogs.
\end{abstract}


\altaffiltext{1}{McWilliams Center for Cosmology, Department of Physics, 
Carnegie Mellon University, 5000 Forbes Ave., Pittsburgh, PA 15213}
\altaffiltext{2}{School of Computer Science, Carnegie Mellon University,
5000 Forbes Ave., Pittsburgh, PA 15213}

\section{Introduction}\label{sec:intro}

As we enter the era of large-scale structure experiments such as LSST,
WFIRST and Euclid, the creation of reliable mock galaxy catalogs will
become increasingly more important. Such catalogs are essential for
correctly characterizing the expected errors in the analyses of these
datasets, calibrating analysis pipelines and ultimately measuring
cosmological parameters (such as the dark energy equation of state)
from galaxy clustering (e.g. \citealt{Aea12}). Making mock catalogs
for different subpopulations of galaxies (e.g. blue versus red, high
$M_{star}$ versus low $M_{star}$, etc.) to study their clustering
properties is also of utmost importance for understanding galaxy
formation and evolution (e.g. \citealt{Cea08,Gea12}). Although these
mock catalogs can be generated relatively quickly using perturbation
theory-based approaches such as that described in \citet{Manera12},
it is well known that these approximations break down at small scales
(e.g. \citealt{CWP09}).

Alternatively, mock catalogs can be created using simulations which
capture non-linear structure growth to smaller scales, and inherently
include redshift-space distortions (velocity information of the
dark matter particles is known). Ideally we would like large-volume
cosmological simulations with both N-body and hydrodynamics, however,
such simulations are computationally expensive. Hence, the large
number of mocks necessary for obtaining robust error measurements
renders this approach impractical. Fortunately, since pure N-body
simulations are relatively inexpensive, the literature has been
rife with methods for populating the dark matter halos found in
these simulations with galaxies. One of the most popular methods for
doing this is using halo occupation distributions (HODs) which is an
analytic model for determining the number of galaxies ($N_{gal}$) that
should form in a halo given its properties (e.g. \citealt{Zea09}). More
recently, subhalo abundance matching (SHAM) has become very prevalent
(e.g. \citealt{CWK07,Vea10}). This method relies on being able to
correctly identify the subhalos within a halo, a very difficult problem in
its own right due to resolution limitations (see e.g. \citealt{Bea11}). It
then assumes that each subhalo contains a single galaxy with a stellar
mass or luminosity that is monotonically related to a subhalo property.

While both methods have been shown to produce mock galaxy catalogs
that match observations reasonably well, it would be progressive to
attempt to eliminate the above mentioned assumptions. The sophisticated
non-parametric regression algorithms that form a subcategory of ``machine
learning'' (ML) are ideal for this purpose. To obtain the mapping from
halos to $N_{gal}$, the only assumptions that these model-independent
algorithms require are that such a relationship exists and that it
is a continuous function of the halo parameters. They then proceed
to construct a model from the data itself and hence do not impose
any pre-supposed relationships onto the data. We note that although
ML algorithms are non-parametric in the sense that we do not need to
assume a known relationship between data parameters, they do often
require us to assume some operational parameters such as how severely
poor predictions are penalized. These, however, can be optimized as
described in \S\ref{sec:svm}.

In addition, the ML-based approaches we propose here, like
HOD-based methods, rely only on the properties of the parent dark
matter halo. Hence, we can circumvent the difficulties in subhalo
identification. Another point worth highlighting is that in principle,
these techniques can be trivially extended to understand how halo
properties map onto different subpopulations of galaxies. This provides
a method for making mock catalogs for these different subpopulations
as well.

ML algorithms do, however, need to be trained on large, accurate datasets
in order for them to learn robust mappings between halo properties and
galaxy properties such as $N_{gal}$: a large-volume N-body plus hydro
simulation with reliable galaxy formation would be ideal. At present,
we have not been able to acquire such a simulation and so we use the
Millennium simulation with semi-analytic galaxy formation for this
study. However, it is conceivable that such a simulation will become
available in the near future which can be used to characterize the
halo-galaxy mapping via the ML techniques discussed in this work.

A final point of interest rests in the observation that most HOD methods
typically use the halo mass as the only parameter in their halo-to-galaxy
mappings. An important topic to study is the sensitivity of $N_{gal}$
to other halo parameters. For example, there have been investigations
hinting that the environment of the halo is also an important factor in
determining how many galaxies will form (e.g. \citealt{Gea05,Cea12}). This
information can be gleaned using ML techniques as well, through performing
a ``feature selection'' which picks out the halo properties that best
predict $N_{gal}$.

In \S~\ref{sec:data} we describe the dataset we use derived from
the Millennium simulations. In \S~\ref{sec:ml} we describe the 2 ML
techniques we employ to learn the mapping between halo properties and
$N_{gal}$. In \S~\ref{sec:results} we describe our results, i.e. how well
our predictions match the actual values from Millennium. This section
also includes a discussion on using ML to make mocks for different
subpopulations of galaxies. We conclude in \S~\ref{sec:theend}.

\section{Dataset} \label{sec:data}

We construct our dataset from halo and galaxy catalogs at $z=0$ derived
from the Millennium simulation \citep{S05,Sea05}. These catalogs are
obtained via querying the Millennium online database \citep{Lea06}. The
halo catalogs were generated using a friends-of-friends (FoF) algorithm
with linking length of 0.2 and the semi-analytic galaxy prescription used
to populate these halos is described in \citet{Cea06,dLea06,dLB07}. The
Millennium simulation is run with $2160^3$ particles in a $500\hMpc$
box. The cosmology employed has $\Omega_m=0.25$, $\Omega_b=0.045$,
$\Omega_\Lambda=0.75$, $h=0.73$, $n_s=1$ and $\sigma_8=0.9$.

To obtain our dataset, we search through the Millennium halo catalog and
extract all primary halos (FoF groups) with mass greater than $10^{12}\Ms$
(at present, we are unlikely to observe anything less massive except
in the local Universe). We then match galaxies from the semi-analytic
catalog to these halos, keeping only the primary galaxies of a halo or
subhalo (i.e. those flagged 0 or 1 in the Millennium database). Hence,
we emerge with a halo catalog listing the following 7 parameters: number
of particles in the halo $N_p$, $M_{200}$, velocity dispersion $\sigma_v$,
maximum circular velocity $v_{max}$, half-mass radius $R_{1/2}$, spin and
number of galaxies in the halo $N_{gal}$. Our goal is to train a machine
learning algorithm to predict $N_{gal}$ using the other 6 halo parameters.

The semi-analytic model used to populate the Millennium halos with
galaxies is dependent on various thresholds (such as for gas accretion
and star formation) that are also evolved through time. This quality
makes Millennium an adequate testing ground for ML applications because
the mapping from halo parameters to $N_{gal}$ is much more complicated
than the straight-forward functions normally employed in methods such
as HOD and SHAM. The complexity of these mappings should be closer to
the level we expect from actual N-body simulations with hydrodynamics.

There are 395,832 halos (with 445,983 total galaxies) in our sample
which we use for our basic tests of the ML algorithms. However, since
the Millennium semi-analytic model provides $b,v,r,i,z$ magnitudes and
stellar masses for the galaxies, we can also use these same halos to
learn the mapping between halo properties and $N_{gal}$ for different
subpopulations of galaxies. We perform tests of the ML algorithms
after splitting the halo sample on colour and stellar mass (a proxy for
luminosity) in \S\ref{sec:colour} and \S\ref{sec:mstar} respectively.

\section{Machine Learning Algorithms}\label{sec:ml}

We test 2 different machine learning (ML) algorithms for predicting
$N_{gal}$ from the halo parameters $N_p$, $M_{200}$, $\sigma_v$,
$v_{max}$, $R_{1/2}$ and spin. The first is a support vector machine
(SVM) and the second is a k-nearest-neighbours (kNN) routine, both
described below. They work by ``learning'' a relationship between a
set of input features $\mathbf{X}$ and the value we're interested in
predicting $Y$. SVM and kNN are both non-parametric in the sense that
we do not need to assume a model as traditional methods for populating
halos with galaxies do. The mapping is constructed using information in
the data itself: the data picks the model best suited to it. In addition,
we avoid the messy problem of subhalo finding which is a required step
in any SHAM-based approach; like HOD-based models, our proposed ML
techniques operate on the parent halo itself.

The learning process is accomplished by ``training'' the algorithm on a
set of training data where $Y$ is known. The learned mapping can then be
applied to a test set of $\mathbf{X}$ values with known $Y$ to verify
its accuracy. If the learned relationship appears robust, it can be
applied to a set of $\mathbf{X}$ with unknown $Y$ to make predictions.

In testing the accuracy of the predicted values, we draw on the
mean-squared-error (MSE) defined as
\begin{equation}
\mathrm{MSE} = \frac{\sum_{i=1}^{N} 
\label{eqn:mse}
(Y_{i,test,true} - Y_{i,test,predicted})^2}{N}
\end{equation}
where $N$ is the number of test data points. This effectively measures
a combination of variance (the scatter in the predicted values) and bias
(how different from truth the predicted values are).

\subsection{Support Vector Machines}\label{sec:svm}

Support Vector Machines (SVMs) work by mapping the features $\mathbf{X}$
to a higher dimensional-space and attempting to separate them into
regions that map onto specific $Y$ values using a set of hyperplanes
\citep{CV95}. In its original form, it is a classification scheme but
can be generalized to a regression algorithm \citep{Dea97}.

The general idea behind SVM is best illustrated through the case of a
binary classifier. The binary SVM is trained on a set of input features
$\mathbf{X} = \{X_1,X_2,X_3...X_d\}$, where $d$ is the total number of
features, to classify data into one of two classes $Y=\{-1,1\}$. Consider
a set of $N$ training data, each with a corresponding column vector of
features $\mathbf{X}_i$ and a corresponding class $Y_i(\mathbf{X}_i)
= \pm 1$. An SVM attempts to separate the training data into their
appropriate classes using 2 parallel hyperplanes in a high-dimensional
space. These planes can be written as
\begin{equation}
\mathbf{W} \cdot \mathbf{X} - b = \pm 1
\end{equation}
where $\mathbf{W}$ is the normal vector to the hyperplane and $b$ is
some constant scalar analogous to a $y$-intercept in 2D. The 2 planes
must satisfy the condition that no points fall in between them, i.e.
\begin{equation}
\mathbf{W} \cdot \mathbf{X}_i - b \geqslant 1
\end{equation}
for $\mathbf{X}_i$ of the first class (i.e. $Y(\mathbf{X}_i)$=1) or
\begin{equation}
\mathbf{W} \cdot \mathbf{X}_i - b \leqslant -1
\label{eqn:hyperplane}
\end{equation}
for $\mathbf{X}_i$ of the second class. Note that this can be simply
re-written as
\begin{equation}
Y_i (\mathbf{W} \cdot \mathbf{X}_i - b) \geqslant 1.
\label{eqn:linsvmcon}
\end{equation}

The region bounded by these 2 planes is known as the margin, which has
width 
\begin{equation}
w = \frac{2}{||\mathbf{W}||}.
\end{equation}
The best classifier is obtained through maximizing this distance
or minimizing $||\mathbf{W}||$ subject to the constraint in Equation
(\ref{eqn:linsvmcon}) since in this limit we will obtain the most robust
separation of the datapoints. For computational purposes, we actually
end up minimizing $\frac{1}{2}||\mathbf{W}||^2$. The optimization can
be performed using Lagrange multipliers ($\alpha_i$). This leads to
minimizing $||\mathbf{W}||$ with respect to the Lagrangian
\begin{equation}
L(\alpha_i) = \frac{1}{2}||\mathbf{W}||^2 - \sum^{N}_{i=1}\alpha_i
[y_i(\mathbf{W} \cdot \mathbf{X}_i - b) - 1]
\label{eqn:lp}
\end{equation}
subject to the constraints $\alpha_i \geq 0$. Taking the derivative of
this Lagrangian with respect to $||\mathbf{W}||$ yields the solution
\begin{equation}
||\mathbf{W}|| = \sum_{i=1}^{N} \alpha_i Y_i \mathbf{X}_i.
\end{equation}
Values for the $\alpha_i$ can be obtained by substituting this solution
back into Equation (\ref{eqn:lp}) which yields
\begin{equation}
L(\alpha_i) = \sum_{i=1}^{N} \alpha_i - \frac{1}{2}\sum^{N}_{i=1}
\sum^{N}_{j=1} \alpha_i \alpha_j Y_i Y_j \mathbf{X}_i \cdot \mathbf{X}_j
\label{eqn:ld}.
\end{equation}
and maximizing $L(\alpha_i)$ with respect to the $\alpha_i$. It turns
out that only a few of the $\alpha_i$ are non-zero. These correspond
to the training points that satisfy the equality condition in Equation
(\ref{eqn:linsvmcon}). Such points are called the support vectors and
set the value of $b$, i.e. ($b = \mathbf{W}\cdot\mathbf{X}_i - Y_i$). In
practice, $b$ is taken to be an average over the support vectors, that is
\begin{equation}
b = \frac{1}{N_{SV}}\sum_{i=1}^{N_{SV}} (\mathbf{W} \cdot \mathbf{X}_i
- Y_i)
\end{equation}
where $N_{SV}$ is the number of support vectors.

It is often useful to introduce some slack (quantified by $\xi_i$ below)
into the SVM classifier. This amounts to replacing the constraint in
Equation (\ref{eqn:linsvmcon}) with the new constraint
\begin{equation}
Y_i (\mathbf{W} \cdot \mathbf{X}_i - b) \geqslant 1 - \xi_i.
\end{equation}
The effective reduction of the margin width in the above equation allows
for some misclassification of the data. We then minimize
\begin{equation}
\frac{1}{2}||\mathbf{W}||^2 + C\sum_{i=1}^{N}\xi_i
\label{eqn:nlsvm}
\end{equation}
where $C$ is a parameter that determines the penalty for any
misclassification. If we solve the Lagrangian for this case, we will
find that $\xi_i = \alpha_i/C$ is required.

In addition, one can see that Equation (\ref{eqn:ld}) contains the term
$\mathbf{X}_i \dot \mathbf{X}_j$ which is just the dot product between
two vectors in feature space. We can then imagine generalizing this dot
product to the dot product in a space spanned by a non-linear mapping
($\Phi$) of the features \citep{ABR64,BGV92}. This mapping can be
used to take the features into a higher dimensional space, making
them more easily separable. The dot product $\Phi(\mathbf{X}_i)
\dot \Phi(\mathbf{X}_j)$ can be thought of as a kernel function
$k(\mathbf{X}_i,\mathbf{X}_j)$. Commonly used kernels like the polynomial,
Gaussian (or radial basis function, rbf) and sigmoid functions are
popular due to their simplicity. These have the forms
\begin{equation}
k_{poly}(\mathbf{X}_i,\mathbf{X}_j) = (\mathbf{X}_i \dot \mathbf{X}_j)^m
\end{equation}
where $m$ is the degree of the polynomial,
\begin{equation}
k_{rbf}(\mathbf{X}_i,\mathbf{X}_j) = \exp(-\gamma ||\mathbf{X}_i - 
\mathbf{X}_j||^2)
\end{equation}
and
\begin{equation}
k_{sigmoid}(\mathbf{X}_i,\mathbf{X}_j) = \tanh(\mathbf{X}_i \cdot 
\mathbf{X}_j + r),
\end{equation}
where $\gamma$ and $r$ are parameters of the kernel. 

The simple binary SVM described above can be generalized to support
vector regression (SVR) \citep{Dea97} which is the algorithm we employ
in this study. For the regression problem, we seek hyperplanes satisfying
the equations
\begin{eqnarray}
Y_i - (\mathbf{W} \cdot \mathbf{X}_i - b) &\leqslant& \epsilon + \xi_i \\
(\mathbf{W} \cdot \mathbf{X}_i - b) - Y_i &\leqslant& \epsilon + \xi_i^*
\end{eqnarray}
Here, $\epsilon$ is a tolerance parameter, i.e. there is no penalty
assigned to predictions that fall within $\epsilon$ of the true
value. $\xi_i>0 $ and $\xi_i^*>0$ are slack variables corresponding to
upper and lower constraints on the system output.

The quantity the SVR must minimize is
\begin{equation}
\frac{1}{2}||\mathbf{W}||^2 + C\sum_{i=1}^{N}(\xi_i+\xi_i^*).
\label{eqn:svr}
\end{equation}
which closely resembles Equation (\ref{eqn:nlsvm}). The Lagrangian takes
the form
\begin{eqnarray}
L(\alpha_i,\alpha_i^*) = && 
\frac{1}{2}||\mathbf{W}||^2 + C\sum_{i=1}^{N}(\xi_i + \xi_i^*) \\ \nonumber
&& - \sum_{i=1}^{N} \alpha_i [\epsilon + \xi_i-y_i + (\mathbf{W} \cdot 
\mathbf{X}_i + b)] \\ \nonumber
&& - \sum_{i=1}^{N} \alpha_i^* [\epsilon + \xi_i^*+y_i - (\mathbf{W} \cdot
\mathbf{X}_i + b)] \\ \nonumber
\end{eqnarray}
where $\alpha_i$ and $\alpha_i*$ are Lagrange multipliers subject to
the constraints $0\leq\alpha_i\leq C$, $0\leq\alpha_i^*\leq C$ and
$\sum_{i=1}^{N} (\alpha_i - \alpha_i^*) = 0$.

For our analysis, we use the SVR algorithm implemented in the scikit-learn
Python library \citep{Pea11}. We use the default $\epsilon=0.1$ and find
that changing this value does not make a noticeable difference in our
predictions or the resulting MSE. The algorithm takes in a value for the
penalty parameter $C$ from Equation (\ref{eqn:svr}) and the kernel. $m$,
$\gamma$ and $r$ need also be specified depending on what kernel is
being used. Optimal parameter values can be determined by splitting our
sample of halos into three equal parts: a training set, a validation set
and a test set. We can then train the SVM using some pre-determined grid
values: $C=\{100,...,10^9\}$ and $\gamma$ or $r=\{0.1,...,10^{-8}\}$ in
powers of 10, and calculate the MSE of the validation set. The parameter
values that give the minimum MSE are chosen for our analyses, where we
evaluate how well the ML predictions match truth using the test set.

\subsection{k-Nearest-Neighbours}\label{sec:knn}

The k-Nearest-Neighbours (kNN) algorithm is much simpler than the
SVM to understand and can also be used for both classification and
regression. The kNN routine calculates ``distances'' to the $k$ nearest
training data points for each point $\mathbf{X}_i$ that we are interested
in predicting $Y_i$. This distance is often just a simple Euclidean
distance between the features $\mathbf{X}$, however, one can imagine
using other definitions as well. The predicted $Y_i$ is then just the
average of the $k$ nearest training set $Y$ values. This average can be
weighted according to distance such that points further away have less
impact on the predicted value. Mathematically, one can represent this as
\begin{equation}
Y_i = \frac{\sum_{k} w(d(\mathbf{X}_k,\mathbf{X}_i))Y_k }{k}
\end{equation}
where $d(\mathbf{X}_k,\mathbf{X}_i)$ is the distance between
$\mathbf{X}_i$ and one of its $k$ nearest neighbours $\mathbf{X}_k$,
and $w(d)$ is the weight corresponding to that distance.

We again use the scikit-learn implementation of kNN \citep{Pea11}. The
algorithm takes in a value for $k$. Again, the optimal value can be
found by stepping through a pre-determined set $k=\{3,6,...,21,24\}$
and picking the value with the lowest MSE when the algorithm is applied
to the validation set.

\subsection{Feature Selection}\label{sec:fs}

Feature selection is the process by which we select which features are the
most relevant for predicting $Y$ from $\mathbf{X}$. A simple approach to
this is forward feature selection which relies on an initial comparison
to the base MSE, or the MSE calculated by taking all $Y_{i,test,predicted}
= \langle Y_{train} \rangle$ in Equation (\ref{eqn:mse}), i.e.
\begin{equation}
\mathrm{Base \; MSE} = \frac{\sum_{i=1}^{N} 
(Y_{i,test,true} - \langle Y_{train} \rangle )^2}{N}
\end{equation} 
This is just the MSE one would obtain by doing the most naive thing:
predicting all $Y$ values to be the mean $Y$ of the training set (denoted
as $\langle Y_{train} \rangle$ above.

Forward feature selection starts by training an ML algorithm to predict
$Y$ using only a single feature. We repeat this for each individual
feature and calculate its MSE value from a test set. If the minimum
MSE is less than the base MSE, then we are doing better than the naive
prediction, indicating that the features do contain information that
is correlated with and can help us predict $Y$. We then ``select''
the feature that produced the minimum MSE and individually add each
other feature to it and repeat the train/test procedure to calculate
MSE values. At the end of this round, if the minimum MSE is smaller than
the minimum MSE in the previous step, we again select the feature that
produced this minimum MSE and repeat the previous procedure. At each step
if adding in an additional feature decreases the minimum MSE further, we
continue. Otherwise we stop and deem the remaining features as not having
much predictive power beyond the ``selected'' features.

Such feature selection schemes are useful for identifying the halo
parameters that are most relevant to inferring $N_{gal}$.

\section{Results}\label{sec:results}

As described in \S~\ref{sec:data}, our tests use a sample of 395,832
halos from the Millennium simulation to assess the machine learning
algorithms detailed in the preceding section. We randomly split this
halo sample into three equal parts and use the first part for training,
the second part for validation and the third part for testing. The base
MSE of the test set is $\sim0.505$. We first look at the results obtained
through training the ML algorithms on all 6 halo features.

As mentioned in \S\ref{sec:svm}, we do a grid-search to find the SVM
training parameters ($C$, $\gamma$ and kernel) that return the minimum
MSE on the validation set. This procedure selected the rbf kernel with
$C=1000$ and $\gamma=0.0001$. We then use the SVM trained using these
values to make predictions from the test set. For kNN, we use a similar
search technique (described in \S\ref{sec:knn}) and find that using
$k=12$ gives the minimum MSE on the validation set. The kNN test set
results below are derived using this value.

\begin{figure}
\vspace{1cm}
\centering
\epsfig{file=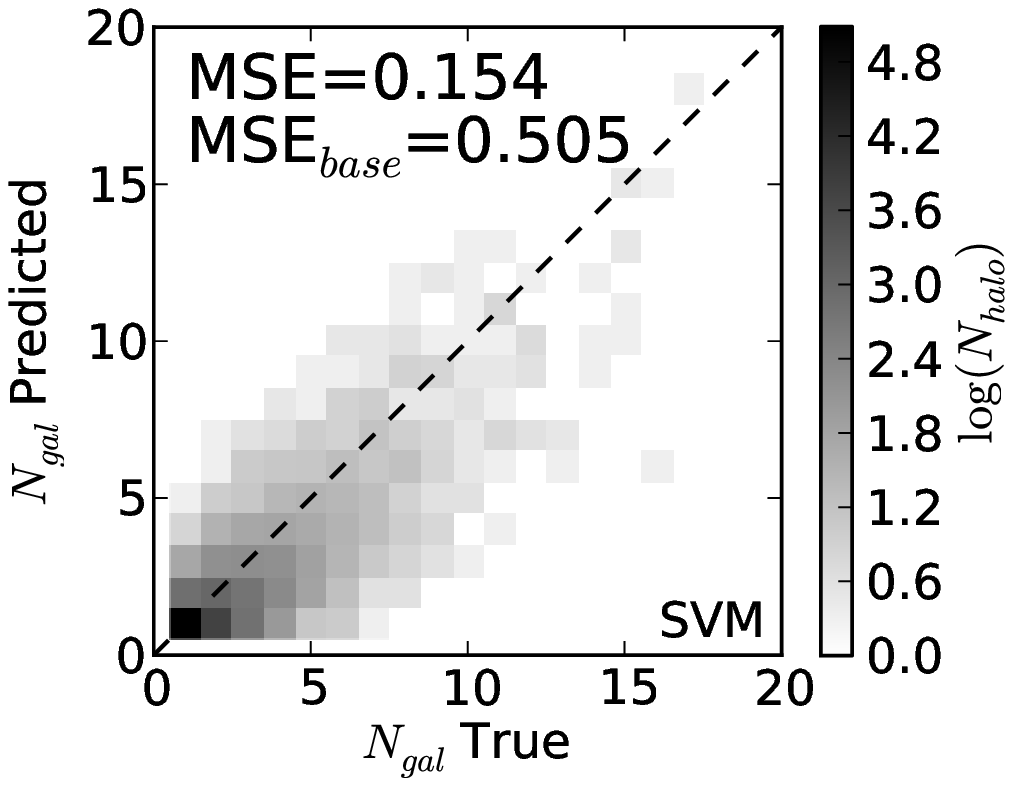,width=0.85\linewidth,clip=}
\epsfig{file=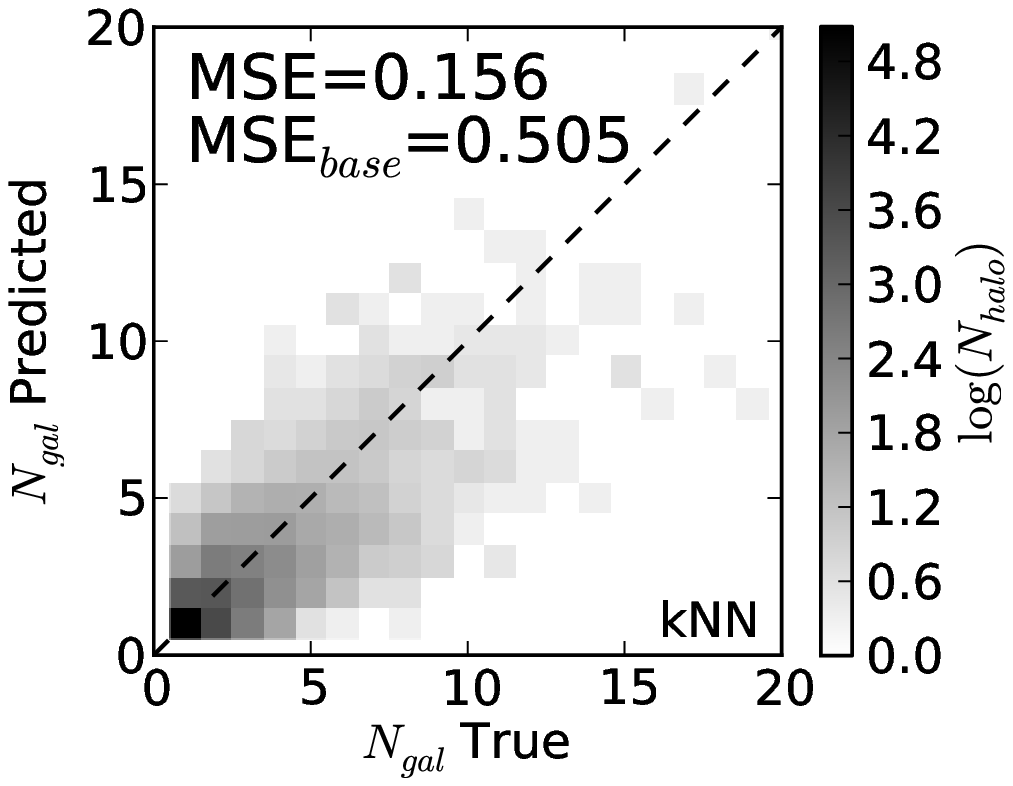,width=0.85\linewidth,clip=}
\caption{2D histograms of machine-learning-predicted number of galaxies
per halo ($N_{gal}$) versus the true number. Here we have taken the
395,832 halos with $M>10^{12}h^{-1}M_{\odot}$ from the Millennium
simulation and split them randomly and equally into training, validation
and test sets. We then use the following features to predict the
mapping from halo properties to $N_{gal}$: number of particles $N_p$,
$M_{200}$, $\sigma_{v}$, maximum circular velocity $v_{max}$, half-mass
radius $R_{1/2}$ and halo spin. The ``goodness'' of the prediction is
indicated by the MSE (see Equation \ref{eqn:mse}) which is essentially
a combined measure of variance and bias. The dashed black line in each
panel indicates the 1:1 line. (top) Results from the support vector
machine (SVM) algorithm. (bottom) Results from the k-nearest-neighbours
(kNN) method. One can see that the MSE is fairly small in both cases. For
comparison, the base MSE (the MSE one would obtain if one always predicted
$N_{gal}$ to be the average of the training set) is $\sim0.505$, which is
a factor of a few larger than the MSE values derived from our ML-predicted
$N_{gal}$. There appears to be a small bias towards under-prediction of
$N_{gal}$, however this does not seem to detract significantly from our
ability to create large-scale structure mocks.
\label{fig:milall}}
\end{figure}

A set of 2D histograms in $N_{gal,true}$ versus $N_{gal,predicted}$
from our test set is shown in Figure \ref{fig:milall}. The top panel
shows the SVM result and the bottom panel shows the kNN result. One can
see that in both cases, the MSE is dramatically improved over the base
MSE which indicates that the ML algorithms are learning some information
about $N_{gal}$ from the input features as expected. The MSE values from
the 2 different methods are very similar and hence SVM and kNN appear
to be equally good for inferring $N_{gal}$ from halo features. However,
we note that upon careful inspection of the 2D histograms, one sees
that there is a slight bias towards under-predicting $N_{gal}$ which is
discussed more below.

\begin{figure}
\vspace{1cm}
\centering
\epsfig{file=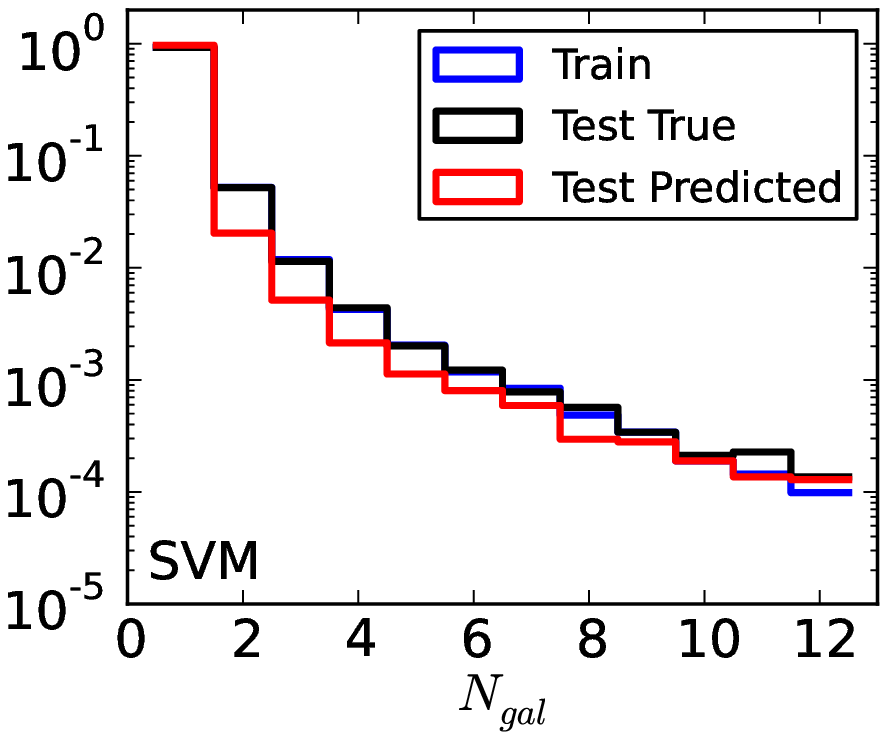,width=0.8\linewidth,clip=}
\epsfig{file=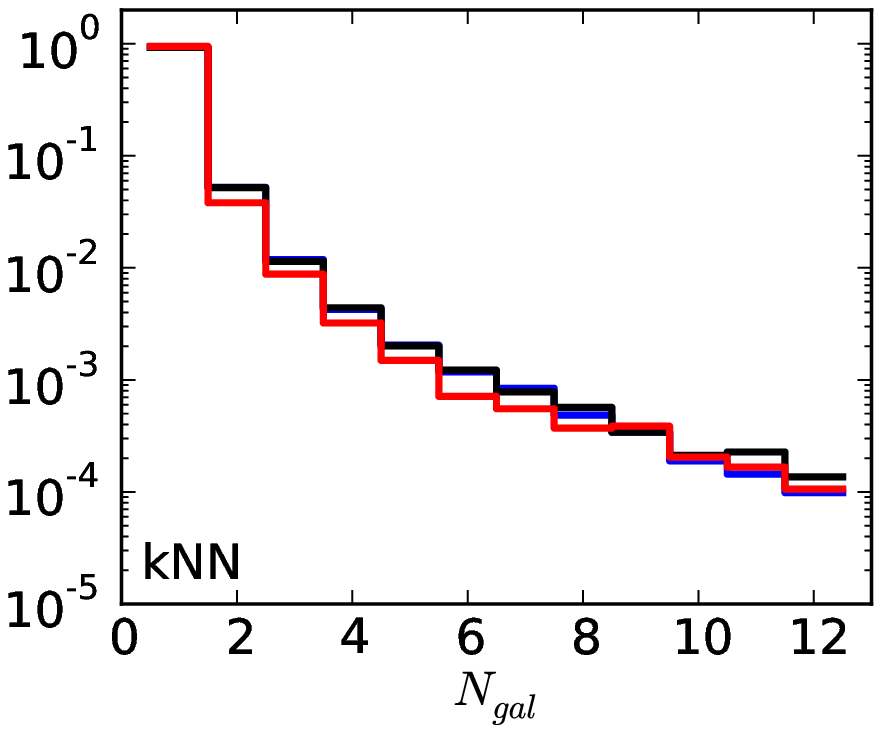,width=0.8\linewidth,clip=}
\caption{Distribution of training, true and predicted test $N_{gal}$
per halo from the Millennium simulation. (top) SVM results. (bottom) kNN
results. One can see that the overall distribution of galaxies predicted
using the ML algorithms track the true values well. It appears that we
slightly overpredict the number of halos with $N_{gal}=1$ and underpredict
elsewhere, however, this does not seem to significantly affect our
ability to make mock catalogs for large-scale structure analyses.
\label{fig:mildist}}
\end{figure}

ML algorithms appear to match the expected distribution of $N_{halo}$ as a
function of $N_{gal}$ quite well as shown in Figure \ref{fig:mildist}. The
panels show histograms where the $y$-axes correspond to the fraction of
halos with $N_{gal}$ and the $x$-axes correspond to $N_{gal}$. The top
panel was obtained using the SVM method and the bottom panel shows the
analogue for kNN. We slightly overpredict the number of halos with 1
galaxy, and underpredict elsewhere, especially when using the SVM. The
true number of galaxies in the test set is 149,064; for SVM we predict
140,519 and for kNN we predict 144,346. This phenomenon was pointed out
previously and will be elaborated on below.

\begin{figure}
\vspace{1cm}
\centering
\epsfig{file=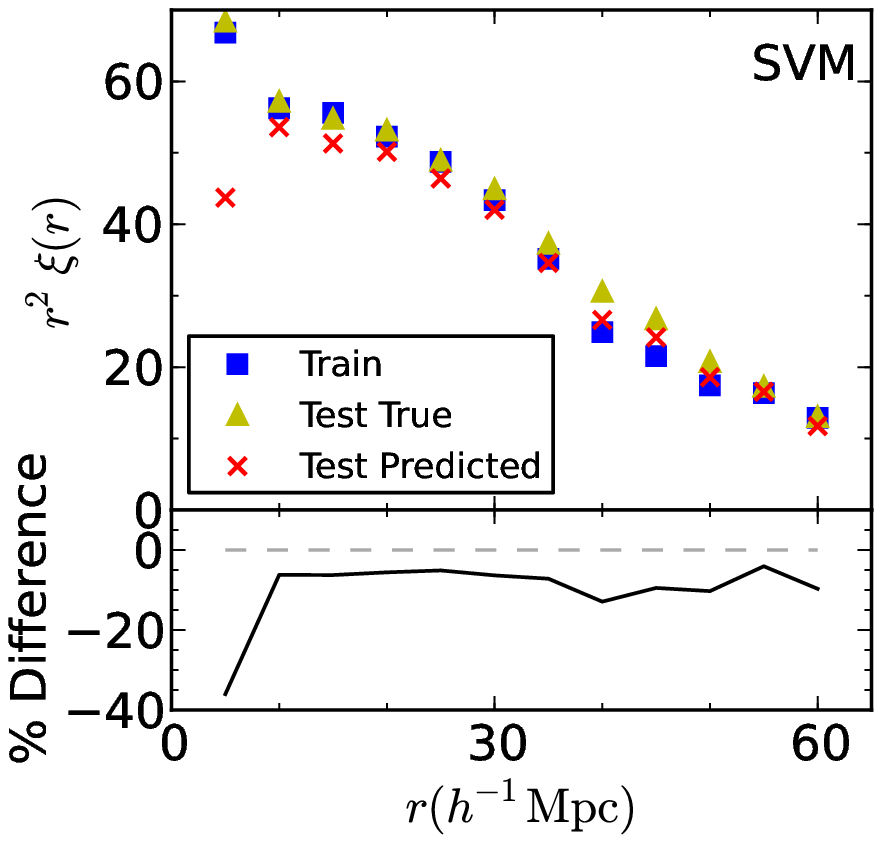,width=0.85\linewidth,clip=}\\
\epsfig{file=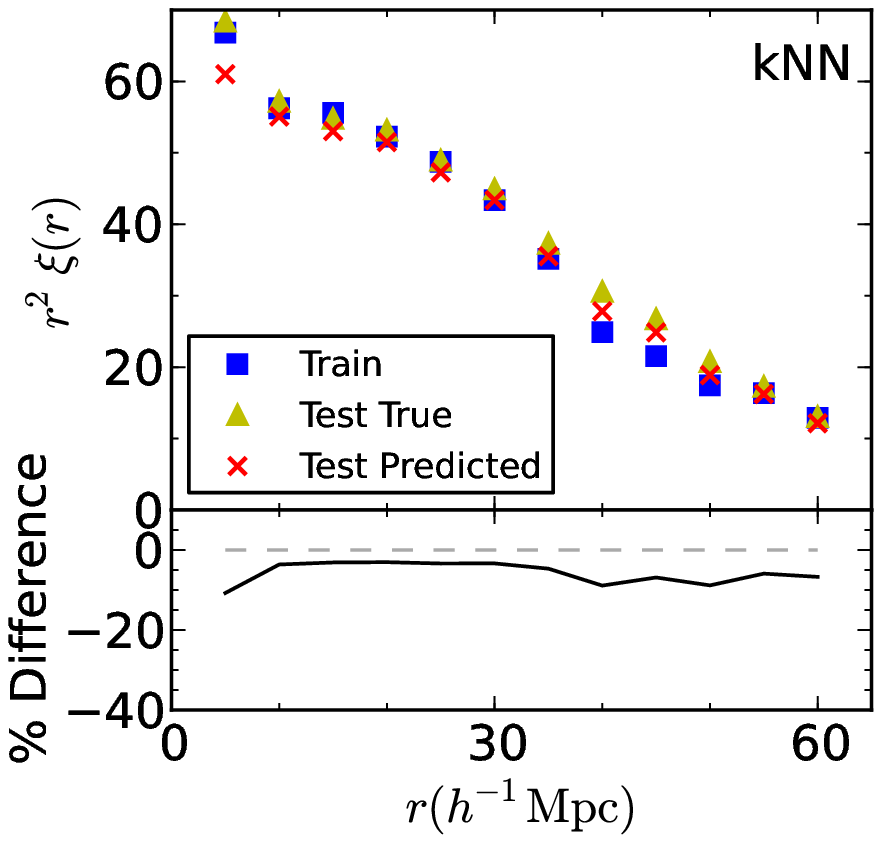,width=0.85\linewidth,clip=}
\caption{Correlation functions $\xi(r)$ of the training, true test set
and predicted test set galaxies. We create a mock galaxy catalog from
the ML-predicted galaxies in order to calculate their $\xi(r)$. This is
an excellent test of the ML predictions as $\xi(r)$ is a key observable
used for large-scale structure analysis, a major motivator for generating
mock galaxy catalogs in the first place. The bottom panels of each
plot show the percentage difference between the $\xi(r)$ calculated
from the actual test set galaxies and the ML-predictions. The grey
dashed line marks 0\% to help guide the eye. The correlation functions
of the ML-predicted galaxies match the true correlation function well
overall. However in the SVM case, due to the underprediction of halos
with $N_{gal}>1$ at small scales (around $5\hMpc$) where the 1-halo
term is still important, the clustering amplitude appears to be 40\%
smaller than expected. This is not a major deterrent from using ML for
making large-scale structure mocks since for these analyses we are mostly
interested in scales greater than $\sim30\hMpc$ where the difference
between true and predicted $\xi(r)$ is mostly $<10\%$. In addition,
the kNN case demonstrates better agreement between predicted and true
$\xi(r)$ at small scales. Hence, ML provides a potentially interesting
alternative for creating large-scale structure mocks.
\label{fig:milxi}}
\end{figure}

The galaxy correlation function $\xi(r)$ can also be used to test the
robustness of the ML results. This test is especially interesting as
$\xi(r)$ is the principal observable for large-scale structure analyses,
the key motivation behind constructing mock galaxy catalogs. We create
a mock galaxy catalog using the $N_{gal}$ values predicted by our ML
algorithms. We place these galaxies randomly within their host halos
according to an NFW profile in the radial direction and random point
generation on a sphere in the angular directions. We calculate the
correlation function for the training, true test set and predicted test
set galaxies in $5\hMpc$ bins in the range $5\hMpc$-$60\hMpc$ (suitable
for redshift-space distortion analyses from large-scale structure). Figure
\ref{fig:milxi} shows the resulting $\xi(r)$ using SVM (top) and kNN
(bottom). Each plot has 2 panels, the upper panel shows the actual
correlation functions and the bottom panel shows the percent difference
(i.e. $100 \times [\xi_{predict}(r)-\xi_{true}(r)]/\xi_{true}(r)$) between
the correlation function calculated from the predicted galaxies in the
test set versus that calculated from the true galaxies in the test set.

In the SVM case we see that at small scales ($\sim5\hMpc$), the predicted
correlation function has a lower amplitude than the true correlation
function by a significant amount ($\sim40\%$). For kNN, the difference
is fairly benign ($\sim10\%$). At these small scales, the 1-halo
term is still important and hence the fact that we underpredict the
number of halos with $N_{gal}>1$ as shown in Figure \ref{fig:mildist}
will result in lower clustering amplitudes than expected. However, for
analyses of large-scale structure, we are more interested in $\xi(r)$
at $r>30\hMpc$ where our predicted $\xi(r)$ is only $\sim5-10\%$ lower
than the true correlation function for the most part. As the amount of
underprediction is fairly constant, one can even imagine implementing a
simple scaling correction for crude applications. Hence, the potential
for making large-scale structure mocks using SVM or kNN remains worthy
of further investigation.

As mentioned above, there is a slight bias towards under-predicting
$N_{gal}$. This is likely due to the fact that the number of halos with
small $N_{gal}$ dominates the overall distribution while halos with high
$N_{gal}$ are much rarer (see Figure \ref{fig:mildist}). Then, taking kNN
as an example, even if a halo should have a large number of galaxies,
its nearest neighbours may still be dominated by $N_{gal}=1$ halos,
which will lead the algorithm to underpredict. The current training set
is clearly incomplete for halos with large $N_{gal}$. A larger training
sample will have a proportionately larger number of these halos so
using a larger training set should be able to partially mitigate this
problem. In addition, recall that the best kernel and parameters ($C$
and $\gamma$) for the SVM and number of nearest neighbours $k$ for the
kNN are selected using the validation set on the basis of minimizing
MSE. Since the MSE is a balanced measure of variance and bias, one can
imagine giving a different weighting to the bias, i.e. penalizing the
MSE more if the bias is high. This can potentially reduce the amount
of underprediction we see, however, we will pay the price of having a
larger scatter in our predictions.

For SVM we underpredict the total number of galaxies by $\sim6\%$
and for kNN we underpredict by $\sim3\%$. These are both small and
do not appear to significantly alter the correlation function shown
in Figure \ref{fig:milxi} at scales relevant to large-scale structure
analyses. At these scales, the correlation function is dominated by the
2-halo term which comes mostly from halos containing a single galaxy. As
discussed above, such halos are the most abundant by far. In addition,
Figure \ref{fig:mildist} indicates that the number of halos we predict
with $N_{gal}=1$ matches the true distribution reasonably well. Hence,
it is not surprising that our predicted $\xi(r)$ is in fair agreement
with the true $\xi(r)$ at large $r$.

\begin{figure*}[h]
\vspace{1cm}
\centering
\epsfig{file=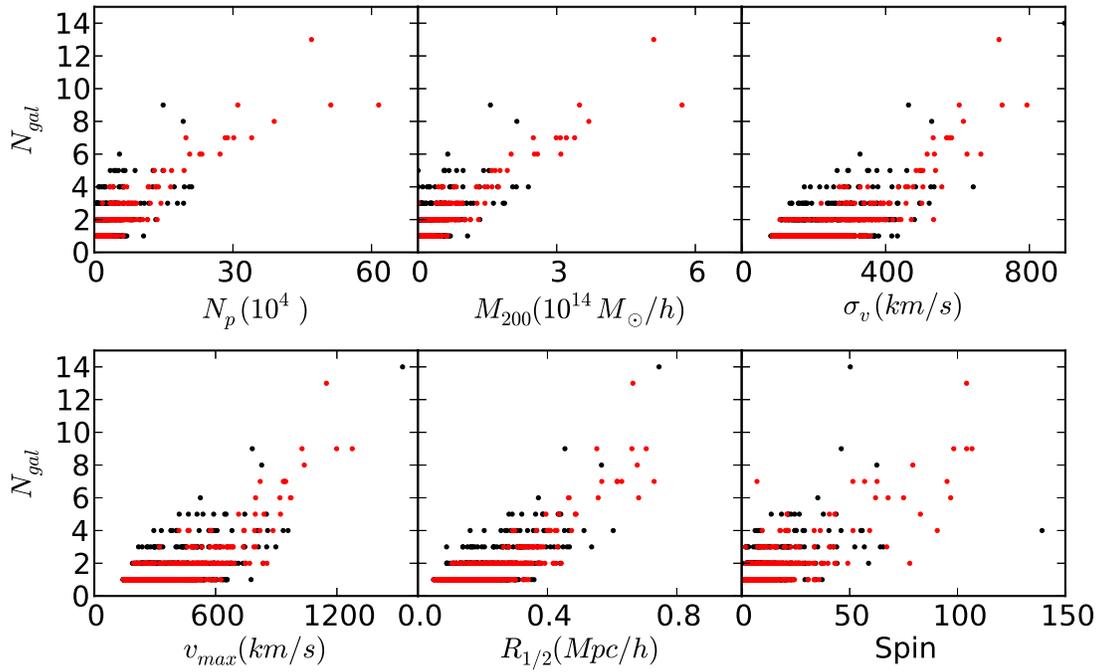,width=0.8\linewidth,clip=}
\caption{Distribution of $N_{gal}$ as a function of the different
features. The black datapoints correspond to the true $N_{gal}$
values from the test set and the red datapoints correspond to the
ML-predicted values. One can see that the true and predicted points
have similar spreads, however, the true points tend to be slightly more
spread out overall. This suggests that the features we have used to
predict the mapping between halo properties and $N_{gal}$ do not fully
capture this relationship, although they do well for the most part. We
should not be surprised though as our features mostly trace halo mass
and environment. In principle, the full merger history of the halos
might impact more than mass and environment. The fact that these 2 key
factors already predict the mapping to $N_{gal}$ as well as shown here
is impressive.
\label{fig:milfeat}}
\end{figure*}

We can also look at the distribution of $N_{gal}$ as a function of the
various features. Figure \ref{fig:milfeat} shows this in scatter plot form
for the kNN test. The analogous plots for SVM are largely the same. Here
we have randomly subsampled the number of points plotted to 3,000. The
black points show true $N_{gal}$ versus features while the red points
show predicted $N_{gal}$. One can see that overall the span of the points
overlaps quite well between the true and predicted sets, again indicating
the general statistical agreement between the two. However, the black
points do show slightly more spread overall. This indicates that the
halo properties we have chosen to use here may not fully encapsulate the
mapping to $N_{gal}$, i.e. we are still missing some information that is
relevant to the halo-galaxy relationship. This should not be surprising
as most of our parameters ($N_p$, $M_{200}$, $\sigma_v$ $v_{max}$)
are effectively mass tracers. The ratio of $R_{1/2}$ and $R_{200}$
(calculated easily from $M_{200}$) can be thought of as a measure
of concentration, effectively the ratio of 2 radii that are defined
in the same way for all halos. Concentration is known to trace mass
\citep{NFW96}, but has also been found to correlate with halo environment
\citep{Wea06,GS07,Mea07}. One can imagine that in addition to mass, spin
and environment, there are additional factors (branching from the full
merger history of the halo) that might affect $N_{gal}$. Fortunately,
it appears that any other factors are not completely orthogonal to the
halo properties used in this study. As shown above, our parameters seem
to capture most of the halo-galaxy mapping, at least in the context of
the Millennium simulations.

Finally, we can perform a feature selection to identify, within
the framework of Millennium, the most predictive halo property for
$N_{gal}$. We employ a forward feature selection algorithm using SVM to
do this. To ensure the stability of our results, we re-perform the feature
selection 10 times with different random draws of the training, validation
and test sets. Key numbers are summarized in Table \ref{tab:fs}. The MSE
values quoted for each feature under the column heading ``first round''
correspond to the median MSE values of the 10 trials in the first round of
the forward feature selection. One can see that $R_{1/2}$ has the smallest
MSE in the first round of selection which indicates that it is the best
predictor for $N_{gal}$. Using $R_{1/2}$ as the only feature for training
gives a median MSE of 0.163 which is very close to the MSE obtained using
all features as shown in Figure \ref{fig:milall}. The other parameters
yield MSE values ranging from 0.189 ($M_{200}$) to 0.275 (spin). The
values quoted under ``second round'' correspond to the median MSE obtained
by adding in another feature on top of $R_{1/2}$. One can see that this
does not significantly change/improve on the minimum MSE from the first
round. Hence it appears that most of our constraint on $N_{gal}$ is coming
from $R_{1/2}$ in the Millennium simulations. The selection of $R_{1/2}$
should not be surprising. It contains information about the halo mass and,
as discussed above, can be related to halo environment through $M_{200}$.

\begin{table}
\centering
\caption{\label{tab:fs}MSE values obtained by performing a forward
feature selection using SVM. The values quoted are the median MSE from 10
different randomizations of the training, validation and test sets. The
halo parameters we use are listed in column 1 and the median MSE values
obtained by using only the listed parameter are shown in column 2
(First Round). One can see that $R_{1/2}$ has the smallest median MSE
and hence it should be the best predictor of $N_{gal}$ in the context
of the Millennium simulations. Adding in additional parameters does not
significantly change the median MSE as indicated in the third column
(Second Round) which lists the median MSE values obtained using $R_{1/2}$
plus the parameter listed in column 1.}
\begin{tabular}{lcc}
\hline
Parameter & First Round & Second Round \\
\hline
$N_p$&0.216&0.170\\
$M_{200}$&0.189&0.163\\
$\sigma_{v}$&0.220&0.163\\
$v_{max}$&0.229&0.163\\
$R_{1/2}$&0.163&--\\
spin&0.275&0.167\\
\hline
\end{tabular}
\end{table}

\subsection{Colour-dependent Mocks} \label{sec:colour}

The Millennium semi-analytic galaxies come with $b,v,r,i,z$ absolute
magnitudes which we can use to define colours and split galaxies into blue
and red subpopulations. This allows us to study whether or not ML-based
methods for learning the halo-galaxy mapping and, most importantly,
making mock catalogs can be directly extended to subpopulations of
galaxies that have different colours. We define blue galaxies to have
$(v-r)<0.7$ and red galaxies to have $(v-r)>0.7$. This gives 175,177
blue galaxies and 270,792 red galaxies.

We again split the 395,832 halos equally into training, validation and
test sets for each case (blue or red) and repeat the previous tests
using kNN. Note that many of our halos now have 0 red or blue galaxies
reducing the size of our effective training sample. In the case of the
blue galaxies, the MSE we obtain after applying kNN is 0.186 as compared
to a base MSE of 0.334. For the red galaxies, we obtain an MSE of 0.293 as
compared to a base MSE of 0.738. In both cases, there is a significant
reduction in MSE compared to the base MSE which indicates that the
algorithm is learning information about $N_{gal}$ from our input features.

\begin{figure}
\vspace{1cm}
\centering
\epsfig{file=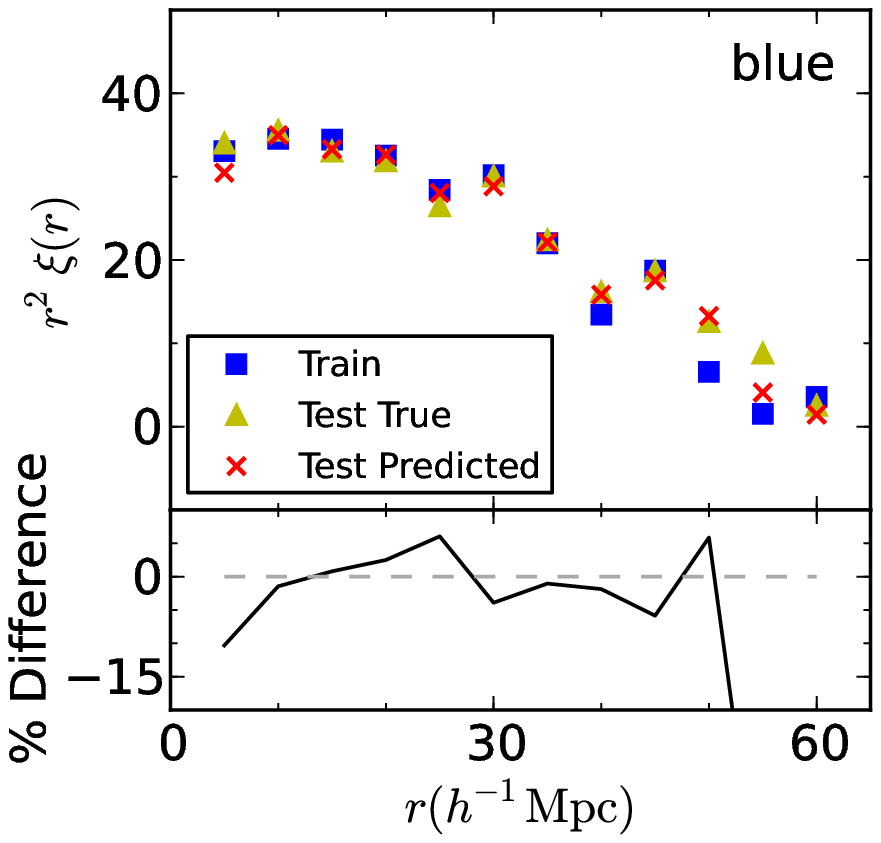,width=0.85\linewidth,clip=}\\
\epsfig{file=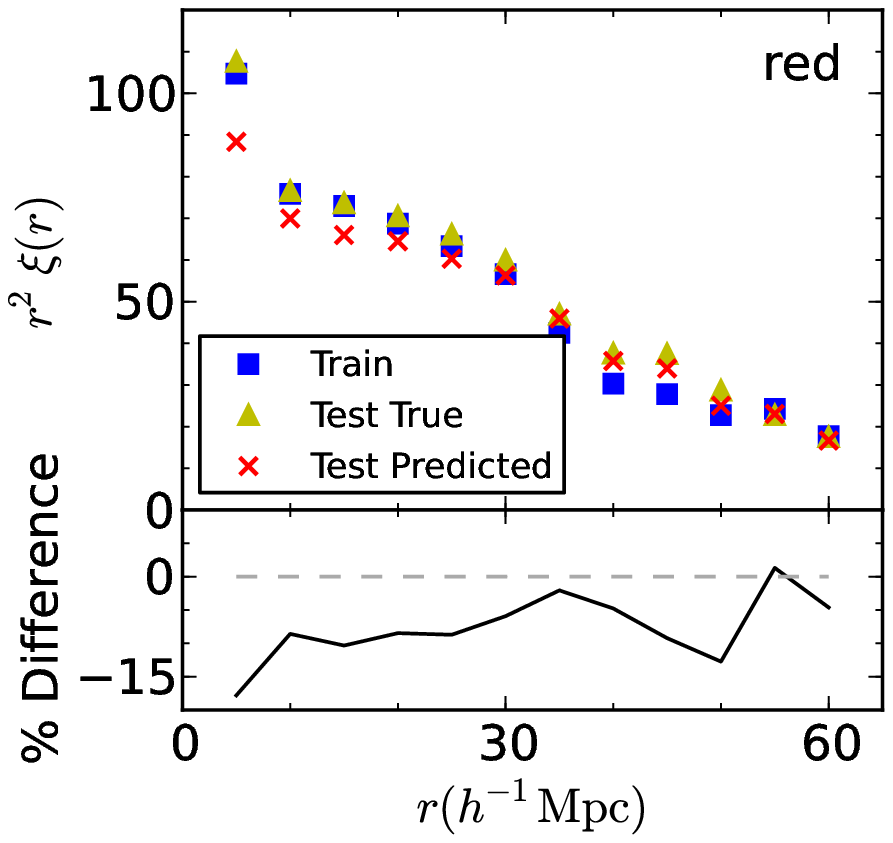,width=0.85\linewidth,clip=}
\caption{Correlation functions of the training, true test set and
predicted test set galaxies. Here, we have separated the galaxies into
blue and red subpopulations. One can see that the general agreement
between predicted and true $\xi(r)$ is not bad with deviations mostly
$\lesssim10\%$. However, there appears to be a slight deficiency in
power at small scales, especially for the red galaxies, in the predicted
$\xi(r)$. This will make the study of galaxy evolution more problematic
since it draws on information contained in the small-scale clustering
of galaxies. However, using ML to make color-dependent mocks at large
scales remains a possibility.
\label{fig:colour}}
\end{figure}

Correlation functions derived from our predictions are shown in Figure
\ref{fig:colour}. One can see that the predicted $\xi(r)$ agrees fairly
well with truth. Again at small scales $\sim5\hMpc$, we see that the
predicted $\xi(r)$ has a lower clustering amplitude, especially for
the red galaxies ($\sim20\%$). If we look at the distribution of halos
with $N_{gal}$, we again see that there is a small under-prediction in
the number of halos with $N_{gal}>1$ that can cause this effect. This is
slightly more problematic here as the main goal of studying the dependence
of clustering on colour is to understand galaxy formation and evolution
mechanisms which rely on information at small scales. However, such
problems can be mitigated if we had a larger set of training data. We are
now also approaching the point where the number of galaxies observed is
large enough for us to begin understanding the differential clustering
of blue and red galaxies at large scales. The agreement between the
correlation function derived from our ML-predicted galaxies and the
true correlation function at these scales is better ($\lesssim 10\%$)
and hence ML-based mock catalogs can be useful for these purposes.

\subsection{Stellar Mass-dependent Mocks} \label{sec:mstar}

Since the Millennium semi-analytic galaxy models also supply us with a
stellar mass, we can split our galaxies into high stellar mass $M_{star}$
and low stellar mass samples. This can help us understand how effectively
we can extend ML-based approaches to understanding the halo-galaxy mapping
as a function of stellar mass, including their potential for constructing
mock catalogs for these distinct subpopulations of galaxies. We make the
cut at $10^{11}M_{\odot}/h$ which gives 71,573 high $M_{star}$ galaxies
and 374,410 low $M_{star}$ galaxies. Stellar mass and luminosity are
correlated with each other and hence by performing this split we are
effectively separating our galaxies into low and high luminosity samples.

Once again we split the halos randomly and equally into training,
validation and test sets. We then run kNN to predict $N_{gal}$ for each
of the high and low stellar mass cases. Note that in the high $M_{star}$
case, most of our halos now contain 0 galaxies so we have reduced the
effective size of the training set by a large amount here. For the high
mass case, we obtain an MSE of 0.119 after applying kNN as compared to
a base MSE of 0.250. For the low mass case, our MSE is 0.214 as compared
to a base MSE of 0.283. Again, the MSE after putting the data through a
kNN algorithm is smaller than the base MSE, indicating that the algorithm
is learning about $N_{gal}$ from the input features.

\begin{figure} 
\vspace{1cm} 
\centering
\epsfig{file=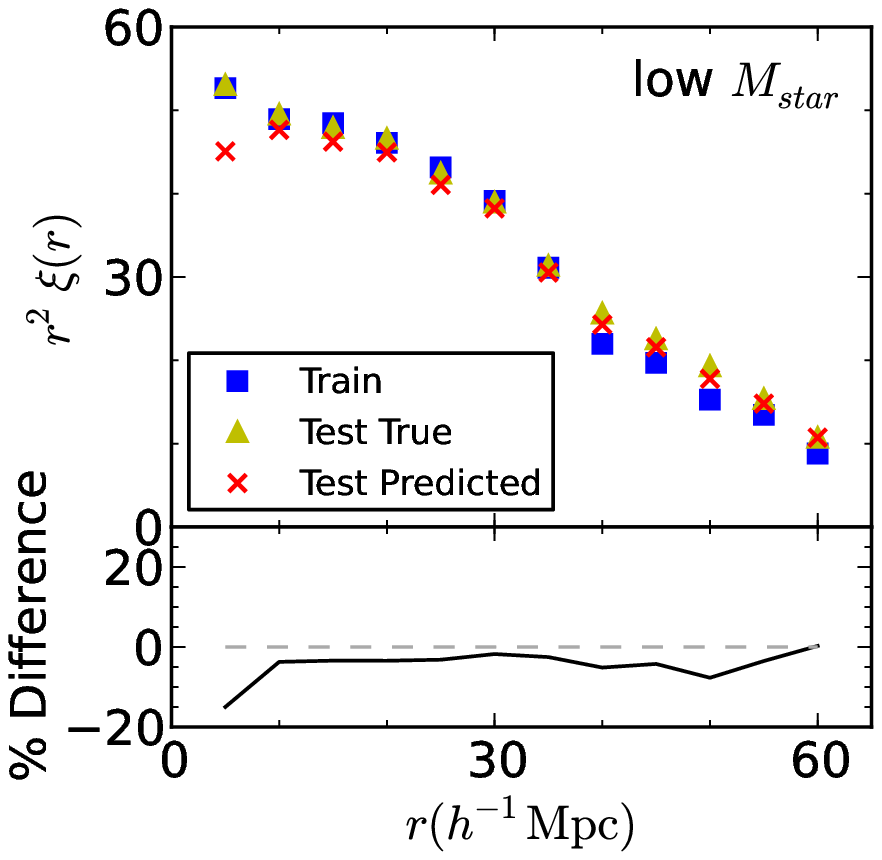,width=0.85\linewidth,clip=}\\
\epsfig{file=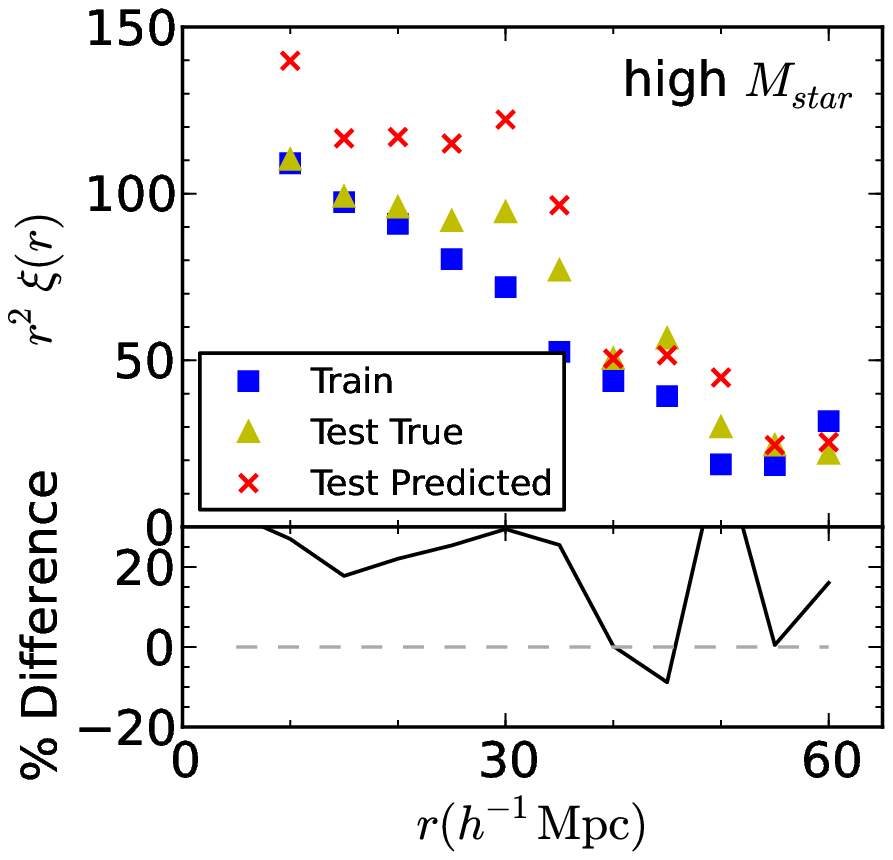,width=0.85\linewidth,clip=} 
\caption{Correlation functions of the training, true test set and
predicted test set galaxies. Here we have split the galaxies into high
and low stellar mass samples. One can see that in the high $M_{star}$
case, the agreement between predicted and true $\xi(r)$ is poor with
$\gtrsim20\%$ differences at most scales. However, this is due to the
fact that high stellar mass galaxies are very rare. Most halos do not
contain any high $M_{star}$ galaxies which reduces the effective size
of our training sample by a large amount. In the low $M_{star}$ case,
the predicted and true $\xi(r)$ agree to $\sim5\%$ at most scales. Again,
the slight underprediction of small scale power will make galaxy evolution
studies difficult. However, the good agreement at large scales suggests
that ML-based approaches for making mock catalogs have potential at
these scales.
\label{fig:mstar}} 
\end{figure}

A plot of the correlation functions derived from the ML-predicted galaxies
is shown in Figure \ref{fig:mstar}. The low $M_{star}$ $\xi(r)$ agrees
well with truth at large scales ($\lesssim 5\%$ deviation), however, it is
again low near $5\hMpc$. Like in the above case where we split by colour,
this is slightly troublesome since studying the luminosity dependence
of clustering is also mostly focused on understanding galaxy evolution
through the clustering at small scales. Nonetheless, mocks produced using
our predicted $N_{gal}$ can still benefit studies of luminosity-dependent
clustering at large scales. The agreement in the high $M_{star}$ case is
poor with deviations $\gtrsim20\%$. However, this is because there are
very few high stellar mass galaxies. With only a small number of halos
with non-zero $N_{gal}$, it is not surprising that the algorithm has some
difficulty with the regression. Looking at the overall distribution of
halos with $N_{gal}$ in this case reveals an underprediction of halos
with $N_{gal}>0$ (including $N_{gal}=1$). Having fewer galaxies that
break the $N_{gal}=1$ threshold effectively increases the galaxy bias
and hence the clustering amplitude due to the 2-halo term which begins
to become important near $\sim5\hMpc$ and is dominant at larger scales.

\section{Conclusions}\label{sec:theend}

We have made some preliminary investigations into using machine learning
techniques to populate dark matter halos from N-body simulations with
galaxies. Since it is very computationally expensive to run cosmological
N-body simulations with hydrodynamics, and perturbation theory approaches
tend to have problems on small-scales (such as treating redshift-space
distortions correctly), machine learning serves as a powerful alternative
for creating large numbers of mock galaxy catalogs. These are a key
ingredient in large-scale structure analyses, a quickly emerging area
with the advent of large galaxy surveys such as LSST, WFIRST and Euclid
which will have effective survey volumes of $\sim10-100h^{-3}\rm{Gpc}^3$.

Most importantly, machine learning brackets a subclass of non-parametric
algorithms which provide a unique way to construct the mapping from
halo properties to galaxies. Unlike other techniques such as HOD,
we do not need to pre-suppose a known model for the halo-galaxy
relationship. The only assumption that must be made is that a function
taking halo properties to number of galaxies per halo does exist and that
it is smooth. It also allows us to circumvent the problems in subhalo
identification which affect SHAM-based approaches.

We test 2 machine learning algorithms, support vector machines and
k-nearest-neighbours, on the halos and semi-analytic galaxies in the
Millennium simulation. We use 6 halo properties: number of particles
$N_p$, $M_{200}$, $\sigma_v$, maximum circular velocity $v_{max}$,
half-mass radius $R_{1/2}$ and halo spin, to characterize the mapping
between halo properties and $N_{gal}$ (the number of galaxies that reside
in the halo). We find that both ML algorithms give mean-squared-errors
of $\sim0.16$ for the predicted $N_{gal}$, which is much smaller than the
base MSE 0.505. The overall distribution of number of halos as a function
of $N_{gal}$ is also matched well by the ML predictions. We use our
ML-predicted galaxies to create a mock galaxy catalog and calculate the
correlation function from it. While in the SVM case we see a deficiency
in clustering amplitude at small scales ($\sim5\hMpc$) by $\sim40\%$,
the predicted correlation function tracks that calculated from the true
Millennium galaxies to $\lesssim10\%$ at the scales most relevant to
large-scale structure analyses. This is very important as large-scale
structure science is the key motivator behind the construction of mock
galaxy catalogs.

Due to the rarity of halos with high $N_{gal}$, we do find that there is
a slight bias towards under-predicting $N_{gal}$. This can lead to the
minor deficit in power at small scales in $\xi(r)$ as mentioned above,
however, this does not appear to significantly deter our ability to
make mock catalogs. As previously stated, we obtain a reasonably good
matching between the predicted and test correlation functions at scales
relevant for large-scale structure studies.

We see that the predicted and true $N_{gal}$ values as a function of the
features are similar in spread. However, for a given $N_{gal}$, the spread
in the features is slightly larger for the true values. This suggests that
the features we have used here do not fully capture the mapping between
halo properties and $N_{gal}$, although, they do come very close. One
should not be surprised by this since our features mostly trace halo
mass and environment. While key, the full merger history of the halos is
likely to impact properties of the halo beyond just mass and environment. 

We also demonstrate a simple feature selection procedure on our halo
properties. Feature selection is merely the process by which we use
machine learning to identify the halo property most predictive in the
mapping to $N_{gal}$. In the context of the Millennium simulations, our
feature selection algorithm identifies $R_{1/2}$ as the most relevant
parameter. This is not terribly surprising as $R_{1/2}$ is germane to
both halo mass and environment.

Finally we investigate direct extensions of our ML algorithms to
understanding the halo-galaxy mapping and making mock catalogs for
various subpopulations of galaxies (i.e. blue, red, low $M_{star}$
and high $M_{star}$). We find that in general the agreement between
the predicted and true correlation functions is fair. However, once
again we observe an underprediction of power at small scales in most
cases. As studies of differential clustering in various subpopulations
are aimed at understanding galaxy formation and evolution which draw on
information contained in the small scales of the correlation function,
this is slightly more problematic here. However, we are entering an era
where the number of observed galaxies is large enough to engage in studies
of differential subpopulation clustering at large scales. Our ML-predicted
$\xi(r)$ matches truth to $\sim5-10\%$ at these scales and hence provides
an interesting alternative for creating mocks for such studies.

The key advantage of ML is that it offers a method of inferring the
halo-galaxy mapping in a model-independent manner. In addition, it is
computationally inexpensive: for example, to train an SVM on $\sim170,000$
points as done in this study takes $\lesssim1$ hour on a single core. If
we can run a large cosmological simulation (N-body plus hydrodynamics)
with galaxy formation calibrated against available observations,
we should be able to use the ML algorithms tested here to learn the
mapping from halo properties to $N_{gal}$ very quickly. We can then
create large sets of mock galaxy catalogs from pure N-body simulations
which are much less computationally expensive than running a large number
of these fully-armed cosmological simulations.

There also exist a number of avenues for future investigation including
the use of ML-based approaches to predict not only $N_{gal}$ but also the
positions and velocities of galaxies within the parent halo. This is much
more complex as it requires predicting a distribution (i.e. multiple
galaxy positions $(x,y,z)$ and velocities $(v_x,v_y,v_z)$) for each
halo. One can also imagine devising methods to mitigate the bias
towards underprediction (i.e. penalizing the MSE more for biased
predictions or giving more weight to high $N_{gal}$ neighbours in a
kNN implementation). These will all aid in our quest to generate mock
catalogs reliably and efficiently.

\acknowledgments
We thank Daniel Eisenstein and Martin White for helpful
discussions. X.X. is supported by a McWilliams Center for Cosmology
Postdoctoral Fellowship made possible by the Bruce and Astrid
McWilliams Center for Cosmology. H.T. is supported in part by NSF
grant AST-1109730. M.N. is supported in part by the M. Hildred Blewett
Fellowship of the American Physical Society, www.aps.org. The Millennium
Simulation databases used in this paper and the web application providing
online access to them were constructed as part of the activities of the
German Astrophysical Virtual Observatory.

\end{document}